\documentclass[twocolumn,prl,superscriptaddress,showpacs]{revtex4}
                                                                          
  \usepackage{times}
    \usepackage{graphicx}
                                          
\begin{document}

\title{Proximity Effect in a Superconductor-Metallofullerene-Superconductor 
Molecular Junction}

\author{A.Yu.~Kasumov} 
\email[Electronic address: ]{kasumov@lps.u-psud.fr}
\affiliation{RIKEN, Hirosawa 2-1, Wako, Saitama 351-0198, Japan}
\affiliation{Institute of Microelectronics Technology and High Purity Materials, RAS, Chernogolovka 142432 Moscow Region, Russia}
\affiliation{Laboratoire de Physique des Solides, Associ\'e au CNRS, B\^atiment 510, Universit\'e Paris-Sud, 91405 Orsay, France}
\author{K.~Tsukagoshi}
\affiliation{RIKEN, Hirosawa 2-1, Wako, Saitama 351-0198, Japan}
\affiliation{PRESTO, JST, Honcho 4-1-8, Kawaguchi, Saitama, Japan}
\author{M.~Kawamura}
\author{T.~Kobayashi}
\affiliation{RIKEN, Hirosawa 2-1, Wako, Saitama 351-0198, Japan}
\author{Y.~Aoyagi}
\affiliation{RIKEN, Hirosawa 2-1, Wako, Saitama 351-0198, Japan}
\affiliation{Department of Information Processing, Tokyo Institute of 
Technology, Nagatsuda 4259, Midori, Yokohama, Kanagawa 226-8502, Japan}
\author{K.~Senba}
\author{T.~Kodama}
\author{H.~Nishikawa}
\author{I.~Ikemoto}
\author{K.~Kikuchi}
\affiliation{Department of Chemistry, Tokyo Metropolitan University, 
Minami-Ohsawa 1-1, Hachioji, Tokyo 192-039, Japan}
\author{V.T. Volkov}
\author{Yu.A. Kasumov}
\affiliation{Institute of Microelectronics Technology and High Purity Materials, RAS, Chernogolovka 142432 Moscow Region, Russia}
\author{R.~Deblock}
\author{S.~Gu\'eron}
\author{H.~Bouchiat}
\affiliation{Laboratoire de Physique des Solides, Associ\'e au CNRS, B\^atiment 510, Universit\'e Paris-Sud, 91405 Orsay, France}

\pacs{74.45.+c, 73.63.-b, 61.48.+c, 81.07.-b}

\begin{abstract}
We report low-temperature transport measurements through  molecules 
of Gd metallofullerenes between  superconducting suspended electrodes. 
The presence and number of molecules in the 2 nm-wide gap between electrodes was 
determined by high resolution transmission electron microscopy. We find  that a 
junction containing a single metallofullerene dimer between superconducting 
electrodes displays signs of proximity-induced superconductivity. In contrast, no proximity effects develops in  junctions containing larger cluster of metallofulerenes. These 
results can be understood by taking into account multiple Andreev reflections, and 
the spin states of the Gd atoms.
\end{abstract}

\maketitle

\newpage
The superconducting proximity effect, where superconducting correlations are induced in a non superconducting ("normal", N) metal in contact with a superconductor, has emerged as one of the most powerful tools to investigate the fundamental characteristics of a mesoscopic sample.
One particularly useful configuration is the SNS junction, where "S" stands for superconducting electrodes, and where "N" has in recent years spanned between mesoscopic metal wires \cite{dubos}, molecular wires such as carbone nanotubes \cite{kasu00,morp,buit}, DNA molecules \cite{kasu01}, or even a single atom in the case of breakjunctions \cite{urbina}. Concurrently, thanks to the fabrication of nanometer-sized gaps, it has been possible to investigate the transport properties of small molecules inserted in these gaps. However so far only normal (gold, mostly) electrodes were used \cite{elmol}. Depending on the transmission of the electrode/molecule contact,  transport proceeded via resonant tunneling through the discrete electronic levels of the molecule \cite{spectro} or through Kondo resonances \cite{kondo}. Proximity induced superconductivity in S-Molecule-S junctions (SMS) has not yet been explored and is expected to give rise to interesting phenomena, especially when spin degrees of freedom are involved. In particular  we expect  the formation of Andreev states to lead to non linearities in the IV characteristics and  reveal  phase coherent transport through the molecular levels when they are strongly coupled to the electrodes. In this Letter we report the investigation of transport through dimers and  small clusters of nanometer size molecules (metallofullerenes Gd@C$_{82}$) in good contact with suspended superconducting electrodes. The suspended character of the device allows the observation by high-resolution transmission electron microscopy (HRTEM) of the very same molecules that have been measured.

\begin{figure}
\begin{center} 
\includegraphics[width=7cm]{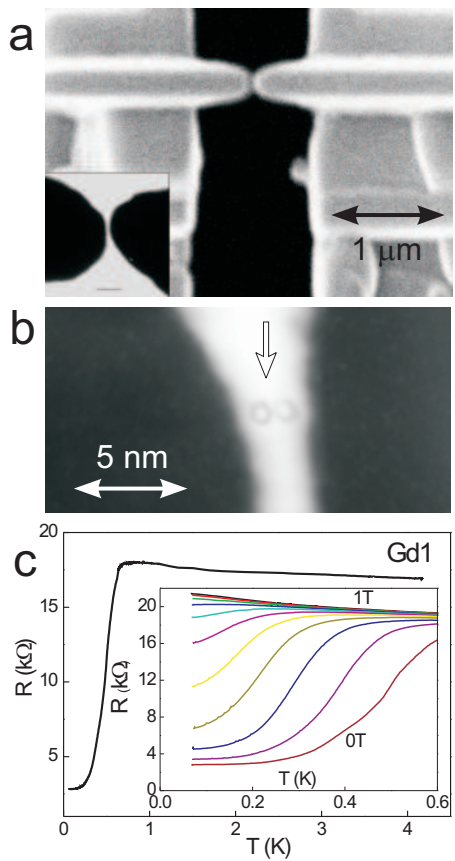}
\end{center}
\caption{Proximity effect in the dimer Gd1. (a) FIB image of W electrodes, suspended above a 
slit cut in a membrane; inset: TEM image of the gap between electrodes: 
scale bar represents 10 nm; rotation of the sample in TEM around the axis parallel to 
the electrodes does not change the gap value. (b) HRTEM image 
of sample Gd1 (Gd@C$_{82}$ molecular dimer between electrodes). 
(c) dc resistances of Gd1 as a function of temperature in zero magnetic field; inset: dc resistance as a function of temperature 
at different magnetic field, perpendicular to current direction. }
\label{fig1} 
\end{figure}
 
Metallofullerene molecules are particularly interesting because the metal atom in the fullerene (Gd here), by acting as a donor (Gd is ionized to state Gd $^3+$), favors charge transfer through the molecule \cite{shino}. Moreover the Gd atoms possess an electronic spin S=7/2 which we find influences the proximity effect through the molecules.

A key experimental achievement  is the design of the molecular junction, which starts with the fabrication and direct 
visualization of a nanometer-size gap between electrodes (fig. 1a), and then enables the trapping, observation and precise identification of molecules in the gap (fig. 1b). This visualization is crucial for proper interpretation of transport measurements, especially to determine the number of molecules within a cluster. In none of previous  
transport experiments  on small molecules \cite{elmol,spectro,kondo} could the molecule be directly visualized.

\begin{figure}
\begin{center} 
\includegraphics[width=7.5cm]{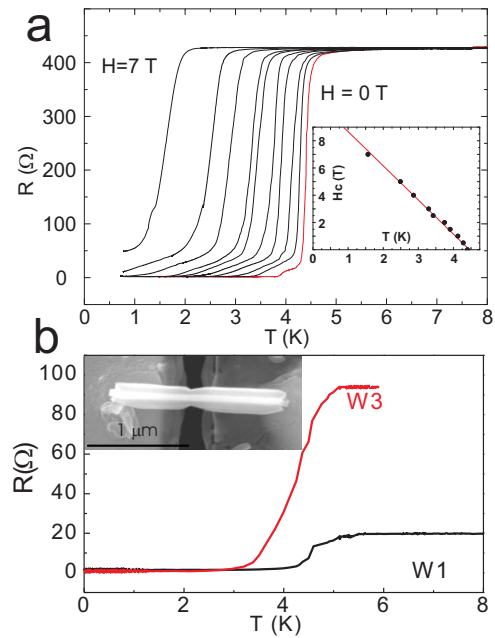}
\end{center}
\caption{Superconducting properties of W 
nanowires, grown by focused ion beam (FIB) decomposition of 
tungsten hexacarbonil. (a) dc resistance of a 200~nm-wide, 
4~$\mu$m-long W nanowire as a function of temperature for 
different values of the magnetic field; inset: Perpendicular critical field 
dependence of the transition temperature (defined as the inflexion point of 
R(T)). Critical current of the nanowire is more than 10~$\mu$A at T=0.7~K.
(b) Resistance versus temperature of the electrodes of Gd1 (W1) and Gd3 (W3), measured after filling in with tungstene the nanogap containing the metallofullerenes. H$_{c}$ is larger than 5 T  below 1 K. Inset: SEM image of W3.}
\label{fig2} 
\end{figure}

We start with a suspended Si$_3$N$_4$ membrane with Au-Ta contacts, through which a micron (or submicron)-wide slit has 
been etched by a focused ion beam (FIB) \cite{kasu00}. We then grow the 200-400 nm-wide, micron-long suspended W nano-electrodes  by local 
decomposition of a tungsten hexacarbonil vapor using a focused Ga ion beam with 
diameter about 5 nm (accelerating voltage 30kV). The growth of these nanowires (at a rate of 0.3~nm/s for the suspended part) is controlled via the display and can be stopped (by switching off the ion 
beam) within a second. It is thus possible to 
fabricate reproducibly two electrodes perpendicular to the slit with a gap 
less than 2 nm wide between them as shown in Fig.~1a. The use of a 
membrane with a slit as a substrate also enables the observation of the gap in a HRTEM. 
The tungsten nanowires grown with this technique are superconducting with a 
transition temperature T$_c$=5~K and a critical field H$_c$ higher than 7~T 
at 0.5~K (Fig.2a). The T$_c$ is that of amorphous tungsten \cite{r17}, 
but H$_c$ is higher because of a large concentration of impurities. 
Auger analysis has shown that FIB-deposited tungsten contains about 10\%~Ga, 
10\%~C and 5\%~O \cite{r18}. 

The deposition of Gd@C$_{82}$ molecules (with purity 99.9\%) 
was carried out as follows. Nitrogen was injected through a capillary 
submersed in a CS$_2$ solution with a 
metallofullerene concentration of 10~$\mu$g/ml (details about molecules in 
solution are given in ref. \cite{r19}). The popping of nitrogen bubbles 
causes microdrops of the solution with molecules to be sputtered at a large 
distance. The sample was placed about 10 mm above the solution, and the 
conduction between the two electrodes was measured under a voltage 
of about 100~mV. When molecules from the microdrops were trapped in the 
gap between electrodes, the resistance dropped from more than 100~M$\Omega$ 
down to a few k$\Omega$, and the nitrogen injection was stopped. We checked 
that the same procedure with CS$_2$ without metallofullerene 
molecules does not cause a resistance drop. It seems that the "desiccation" 
of CS$_2$ microdrops with metallofullerene molecules concentrates 
molecules near and inside the gap. This was confirmed by HRTEM observations 
using a JEOL JEM-2000FX microscope operating at 120~kV. The surprisingly low 
values of  junction resistances indicate good electron transmission 
between the molecules and the electrodes, as in the experiment of ref. \cite{kondo}
(and as predicted by the theory \cite{r20a}).

In the following we report the low temperature transport properties of three samples, Gd1,  Gd2, and Gd3. The HRTEM observations were carried out after the transport 
measurements. Sample Gd1 contains a single Gd@C$_{82}$ dimer 
in the nanogap (Fig.~1b). The 
formation of dimers is known to occur in CS$_2$ solution \cite{r19}, with a 
binding energy of more than 100~meV \cite{shino}, so that the dimer is
stable at a room temperature. According to reference \cite{shino}
the Gd atoms are placed 
asymmetrically with respect to the center of the dimer. Samples Gd2 and Gd3 contain a cluster of Gd@C$_{82}$ molecules between the electrodes (about seven molecules for both samples; contamination during HRTEM observation prevented the exact
determination of this number).

Conductivity measurements of these molecular junctions were carried out in a 
dilution refrigerator at temperatures down to 60~mK, in a magnetic field up 
to 5~T with a nA ac current at 30~Hz superimposed to a dc current varying between
-200 nA to 200~nA. The differential resistance is measured 
 using a  low noise voltage amplifier followed by lock-in detection. 
Sample Gd1, Gd2, Gd3 have room temperature resistances respectively of 13~k$\Omega$, 3~k$\Omega$ and
 1.5~k$\Omega$ and exhibit ohmic behavior for current excitations between 1~nA and 1~$\mu$A.  There is no sign of Coulomb blockade or resonant tunneling in these samples  down to 1~K, certainly due to the good coupling to the W electrodes. Below 1~K, the three samples behave quite differently (Fig. 1, 3, 4).

Sample Gd1 undergoes a transition to a low resistance state 
below 0.7~K (Fig.~1c), much below the 5K transition temperature of the electrodes. The transition is suppressed by a magnetic field of 
1~T. This indicates a transition to a proximity induced superconducting state. The transition temperature and critical field are well 
below the values of the contacts. Note that the transition is not complete 
(no zero resistance state and no supercurrent). 

The physics of electronic transport in Gd1 is rather complex since it is determined by several parameters: the transparencies of the fullerene-electrodes contact, the electronic transfer between fullerenes, and finally the orientation of Gd magnetic moments.
The rather low value of the overall resistance indicates that the transparencies of both the fullerene-electrodes  and the inter fullerene contacts are close to one. The electronic transfer between fullerenes has indeed been shown to be of the order of 50 meV in fullerene crystals doped with alkali metals \cite{dressel}. The level spacing within one fullerene atom is in the 0.1 eV range.
 One can thus consider the dimer as a single quantum dot which level spacing  is much larger than the energy scales involved in the experiment (temperature and voltage drop through the sample). The magnetic coupling between the S=7/2 Gd spins is not precisely known. Magnetisation  experiments performed on powder containing mostly dimers down to 3 K \cite{funa}  exhibit paramagnetism between room temperature and 3K. An average   antiferromagnetic coupling of the order of $J\simeq 0.7 K$,  can be deduced by extrapolating the observed Curie Weiss law describing the susceptibility above 3 K. This antiferromagnetic coupling, larger than the dipolar magnetic coupling energy (which can be estimated to be of the order of 0.1 K), is probably determined by exchange interactions mediated by the 3 Gd atom electrons transfered to the fullerene cage. At low temperature, the dimer is thus expected to be in a frozen non magnetic state where the 2 Gd spins are maintained antiparallel by $J\simeq 0.7 K$. The  most probable cause of  the observed suppression of proximity effect by a rather low  temperature of the order of J and magnetic field $J/gS\mu_B$, is  thus the transition from a non magnetic  
antiparallel to either a fluctuating or a parallel state of Gd atom spins within the dimer \cite{r20}. This will be discussed again below.

The differential resistance (dV/dI)  versus  current was measured for 
different magnetic fields (Fig.~3a) and temperatures (Fig.~3b).  Deducing the
superconducting gap of the W electrodes from the BCS relation 
2$\Delta$=3.52kT$_c$ \cite{r21}, using T$_c$=5~K (Fig.~2b), yields $\Delta$=0.80~meV. It is then possible to estimate a 
critical current I$_c$ from the Ambegaokar-Baratoff formula: 
$I_c=\pi\Delta/2eR_n\sim$~72~nA, that is 
approximately equal to the value corresponding to the first peak in dV/dI 
(Fig.~3c). Note however that we do not observe 
any Josephson current. Also, I$_c$(T) has a non monotonic behavior, with a 
maximum at T=150~mK (Fig.~3c). Such behavior can not be explained by simple 
BCS theory, and may be related to the antiferromagnetic coupling of 
the Gd spins\cite{r22}.

\begin{figure}
\begin{center} 
\includegraphics[width=9cm]{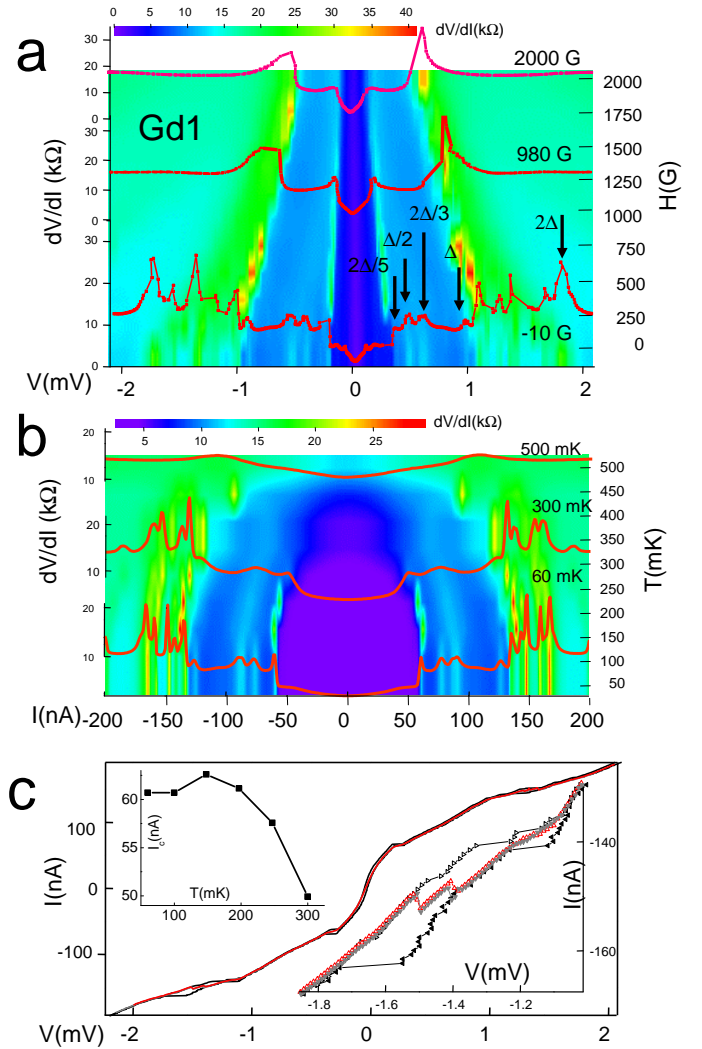}
\end{center}
\caption{(a) Color plot of the field dependence of the differential resistance versus voltage , measured with a small ac modulation of the current superimposed to the dc current  at T=60~mK (data is taken when increasing current from negative to positive values). Arrows indicate submultiple values of the highest bias peak assumed to be at 2$\Delta$; (b) Color plot of the temperature dependence of the differential resistance as a function of current (data is taken when decreasing current from positive to negative values). (c) Comparison between  voltage bias  and current bias data, at 60 mK. Lower right inset: detail of a hysteretic part of the V(I) curve. Upper left inset: Reentrant behavior of the critical current's  temperature dependence.}

\label{fig3} 
\end{figure}

Beside the main peak associated to I$_c$, the differential resistance exhibits 
a complex structure with numerous  hysteretic peaks sometimes not symmetric with respect 
to current reversal.  This hysteresis is not present when the sample is voltage biased. The current is then a non-monotonic function of voltage in the regions where hysteresis takes place in the current-biased data, as observed in Josephson junctions (see Fig. 3c). In long molecular wires between S electrodes, peaks in the differential 
resistance can be attributed to the nucleation of phase slip centers \cite{r23}, but 
there is no room for nucleation centers in short molecules. Rather, we attribute these peaks to multiple Andreev reflection (MAR) \cite{r24} already observed in other SNS junctions \cite{buit,urbina,r25} when the bias is equal to 2$\Delta$/ne, where n is an integer. As shown in Fig. 3a where the differential resistance is plotted as a function of bias voltage, these peaks shift linearly to lower bias with magnetic field following the  field dependence of the gap expected from the transition temperature of the plain tungsten wires depicted in Fig.2.

  However, in addition to the peaks predicted by 
theory \cite{r24}, we also find peaks which the simple MAR model does not 
explain (Fig.~3a). In particular it seems as though the peaks at $2\Delta$ and  $2\Delta/3$ are splitted. This behavior is indeed expected for quantum dots between S electrodes, which contain an energy level not exactly centered at the Fermi energy of the electrodes  \cite{r26}. The amplitude and shape of the peaks have been shown to depend on the transmission of the potential barrier between the superconducting and normal parts of SNS junctions \cite{r24}.   

\begin{figure}
\begin{center} 
\includegraphics[width=7cm]{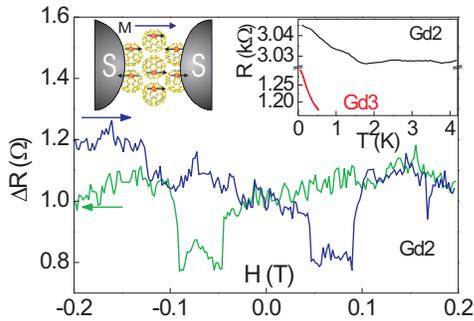}
\end{center}
\caption{Resistance hysteresis for sample Gd2 (molecular cluster) in 
perpendicular magnetic field at T=4.2~K; scan speed is 0.2~mT/s with 10~s wait 
between points. Left inset: schematic representation of the cluster's
 magnetic moment. Right inset: temperature dependence of dc resistance for Gd2 and Gd3 in zero field.}
\label{fig4} 
\end{figure}

The other samples, Gd2 and Gd3, have lower room temperature resistances, indicating a better transmission which should favor proximity effect. Surprisingly they do not undergo a transition to a low 
resistance state, in spite of the fact that a transition was clearly observed in the W electrodes (see Fig. 2). They exhibit a small resistance increase at low temperature (inset of Fig.~4) and a nearly bias independent differential conductance (not shown). The HRTEM observation (not shown) reveals that Gd2 and Gd3 are composed of a cluster of seven or 
more fullerene molecules. We expect such a cluster with more than one dimer to have an uncompensated  
magnetic moment, just like a frustrated antiferromagnetic nanograin. This is because Gd atoms are off-centered within the fullerene cage and can therefore be closer to one another in a cluster, compared to a single dimer, and  thus interact more strongly (Inset of fig.~4). We conjecture that the magnetic moment is the cause of the absence of superconductivity in these junctions. The existence of a magnetic moment  on Gd2 is confirmed by the observation in the magnetoresistance measurements of hysteretic jumps in the 0.1-0.2~T range (Fig.~4). Such behavior is 
characteristic of a ferrimagnetic nanograin, with four  possible values of  the magnetic moment ($\pm M_1$ and $\pm M_2$)  along the field axis, with $M_2 >M_1$. The observation of this hysteretical magnetoresistance was possible at temperatures up to 5K which implies that magnetic order within the cluster takes place at temperatures larger than  the anti ferromagnetic coupling within a single dimer.  Note  also that the highest  moment state ($\pm M_2$) corresponds to higher resistance than the smallest moment one $\pm M_1$. This unusual  behavior could  be due to the superconductivity in the electrodes.  

In conclusion, we have performed the first study of proximity 
induced superconductivity in nanometer-size molecules, along with their observation. This was made possible by the controlled realization of nanometer-size gaps between suspended 
superconducting contacts and the deposition of molecules in the gaps. We find that proximity induced superconductivity is very sensitive to the magnetic state of  the molecules. These experiments  performed with metallofulerenes can \textit{a priori} be transposed 
 to a wide variety of molecules. They can also be used for 
realization of various quantum computation schemes, based on the control of 
a state of quantum dots in contact with superconducting electrodes (for 
example, spin \cite{r28,r29} or mechanical states \cite{r30}).

We thank H. Kataura, Y. Iwasa, S. Okubo, T. Kimura, Y. Otani and A. Furusaki 
for discussions and help. We thank the Russian Foundations for Basic 
Research and Solid State Nanostructures  and Grant-In-Aid for Scientific Research (N 16GS50219) for financial support.


\end{document}